\documentstyle[preprint,aps]{revtex}

\begin{document}
\title{On Cellular Automata Models of Single Lane Traffic}
\draft
\author{M\'arton Sasv\'ari$^{(1,2)}$ and J\'anos Kert\'esz$^{(2)}$}

\address{$^{(1)}$ Institute of Physics, E\"otv\"os University, H-1088 Puskin u.
5-7, Hungary \\
$^{(2)}$ Department of Theoretical Physics, Technical University of Budapest,
Budafoki \'ut 8, H-1111 Hungary}

\maketitle 

\begin{abstract}
The jamming transition in the stochastic cellular automaton model
(Nagel-Schrecken\-berg model) of highway traffic is analyzed in detail,
by studying the relaxation time, a mapping to surface growth problems
and the investigation of correlation functions. Three different
classes of behavior can be distinguished depending on the speed limit
$v_{max}$. For $v_{max} = 1$ the model is closely related to KPZ class
of surface growth. For $1<v_{max} < \infty$ the relaxation time has a
well defined peak at a density of cars $\rho$ somewhat lower than
position of the maximum in the fundamental diagram: This density can
be identified with the jamming point. At the jamming point the
properties of the correlations also change significantly. In the
$v_{max}=\infty $ limit the model undergoes a first order transition
at $\rho \to 0$. It seems that in the relevant cases $1<v_{max} <
\infty$ the jamming transition is under the influence of second order phase
transition in the deterministic model and of the first order
transition at $v_{max}=\infty $. 
\end{abstract}
\pacs{}


\section{Introduction}

The breakthrough in statistical physics due to the understanding of
critical phenomena and the related development of methods has
initiated vivid interdisciplinary research activity. The successful
application of these methods demonstrates that the validity of the
concepts to handle the problem of many interacting units reaches far
beyond the traditional scope of statistical physics.

Vehicular traffic represents, from the point of view of statistical
physics, a far from equilibrium driven system where a combination of
the Highway Code and individual driving strategies replace the usual
physical interactions between particles.  Of course, the approach of a
physicist is quite different from that of a traffic engineer: We would
like to model the typical behavior as simple as possible while
the
essence of some phenomena should remain unaltered though it is not our
aim to describe specific traffic situations.  Nevertheless, one can
hope a twofold gain from these type of studies: First, they could help
us to understand far from equilibrium systems and, second, we strongly
believe that the concepts developed in physics can contribute to the
understanding of the complex phenomena related to vehicular traffic.
Obvious analogies to problems of physics like to kinetic theory or
granular flow have motivated physicists to work in this field.  In
fact, natural scientists' contribution to the understanding of traffic
problems has already a long history \cite{Prig,WSch}.

Numerous articles have been published in the last couple of years
investigating discrete models of highway traffic flow 
\cite{NSch,CsV,N,SSNI,NP}. It was
suggested, that very simple probabilistic models based on cellular
automata can reproduce features of real traffic, including a supposed
transition from low density laminar flow to a high density
phase, where start-stop waves are dominant. The behavior of these
simple models is very complex near this transition, and up to now, is
still not well understood.

The paper is organized as follows. In the next section we summarize
the model and the method of simulation. In Section \ref{relax} the relaxation
time analysis is presented. Section \ref{surface} deals with the relation of 
the traffic models to surface growth. The correlation functions are
analyzed in Section \ref{corr}. The special case of no speed limit is discussed
in Section \ref{vinf}. The paper is finished with a conclusion.

\section{Model and Simulation}

In this paper we consider the cellular automata introduced by Nagel
and Schreckenberg to describe
single lane traffic \cite{NSch}.
The model consists
of a one-dimensional array of $L$ cells with periodic boundary
conditions.
Every cell has $(v_{max}+2)$ states: it can be empty, or it can contain
a car with velocity $v=0,1,\dots,v_{max}$. The density of cars is
$\rho$. We perform the following
steps in parallel for all cars:

{\parindent = 20pt
\parskip=0truecm
---Acceleration: increase $v$ by $1$ if possible.\hfill (1'a)

---Deceleration: decrease $v$ to avoid crash with the car in
front. \hfill (1'b)

---Randomization: decrease $v$ by $1$ with probability $p$ if
possible.\hfill(1'c)

---Movement: move forward $v$ sites. \hfill (1'd)

}

Despite its simplicity this model captures several aspects of highway
traffic including the free flow -- jamming transition.  The parameter
$v_{max}$ can be considered as a limit speed and a usual value for it is
$5$.  We would like to stress the importance of the third step.  The
fact that the model uses {\it braking noise} characterized by the
probability $p$ is crucial.  One could equally introduce random
accelerations, but it can be shown, that these type of perturbations die
out very quickly, see \cite{L}. An initial condition is needed to
specify the model completely.

A convenient way to investigate the model is to draw a diagram of flow
versus density, the so-called {\it fundamental diagram}.  It is a
curve with a well defined maximum at a density $\rho_c$. The occurrence
of density waves is related to the non--linearity of the fundamental
diagram and it is expected, that the jamming transition will occur
somewhere near the maximum.  At low densities the flow is ``free'' with
very few waves due to fluctuations (1'c), which die out
quickly, at high densities above the maximum start--stop waves dominate
the system; it is in the ``jammed'' state.

What is the nature of this transition?  It has been suggested
\cite{NR} that, with increasing density, the jammed regions grow in
space and time and at the jamming point they form a $1+1$ dimensional
interconnected infinite network.  This would point to a percolation
type picture for the process.  However, the percolation transition
depends on the geometry of the clusters, i.e., on the microscopic
definition of jams which introduces some ambiguity.  Moreover, it was
demonstrated in \cite{CsK} that appropriate choice of the parameters
leads, for any reasonable jam definition, to a percolation transition
at densities much higher than the region where jams become dominant.
In order to resolve this problem and to avoid the ambiguity related to
the definition of the jams, in \cite{CsK} another definition of the
jamming transition point was given without explicitly referring to the
density waves.  It was found that relaxation time of the average
velocity has a maximum at a well defined density $\rho_p$ and the value
of the maximum increases with growing system sizes.  This phenomenon
was interpreted in \cite{CsK} as critical slowing down and the critical
point in the infinite size limit was identified with the jamming
transition which turned out to be somewhat below the density at the
maximum in the fundamental diagram ($\rho_c>\rho_p$).

The interpretation of the jamming transition as a second order
non-equilibrium phase transition rises the question of the order
parameter. One suggestion has been \cite{VS} to take the density of jammed
cars as order parameter which shows a rapid increase numerically at
the same density where the relaxation time has its maximum. Besides
the above mentioned problem with the definition of the jam this "order
parameter" has the disadvantage that, due to the random braking events,
there are small jams for any nonzero car density, i.e., the order
parameter would not entirely vanish even below the transition.

The present paper is devoted to a detailed investigation of the jamming
transition. We have studied the model with three different values of
$v_{max}$ namely $1,2$ and $\infty$. Because of the boundary conditions the 
$v_{max}=\infty $ case means $v_{max}=L$. It is widely believed that the 
$v_{max}=1$ and the $1<v_{max}<\infty$ cases differ qualitatively [6] and we
have found that the case without speed limit shows interesting and
unusual features.

With the upper choice of the $v_{max}$ parameters we can compare the three 
classes of models. 

The stationary state of the $v_{max}=1$ model is analytically exactly
solvable \cite{SSNI}. In addition the model has a car-hole symmetry
which connects the stationary state at $\rho$ with the stationary
state at $(1-\rho)$.  In the $v_{max}>1$ cases this symmetry does not
exist because every $v>1$ step corresponds to the common movement of
several holes.  Due to its relationship with other non-equilibrium
models the $v_{max}=1$ model is not expected to exhibit a "phase
transition" at a specific density while this is assumed for
$v_{max}>1$.

For this reason we made simulations for the different cases and methods.
We used $p=0.5$ for the braking probability throughout the paper.
For the $v_{max}=1$ and the $v_{max}=\infty$ cases we used an algorithm
with position and velocity coding, storing the velocities and positions of
the cars in two arrays.
For the $v_{max}=2$ case we used a multi-spin coding algorithm where we
stored the lane $v_{max}+1$ times having $L$ bits for every velocity according
to the system size. In the lane labeled by $v$ the $i$-th bit is $1$ when 
there is a car with velocity $v$ at site $i$ and bit zero otherwise.
This algorithm is faster because we can use bitwise operations.
When calculating velocity correlation functions, we used the multi-spin 
coding algorithm only for reaching the stationary state and then switched
to the position coding technique what was more appropriate for the 
calculations afterward.
 
\section{Relaxation time analysis}
\label{relax} 
First we measured the average car velocity $\overline{v}={1 \over N}
 \sum_i v_i$ as a function of time. As an initial condition we used
 uniformly distributed cars with $v_i=0$ for all cars and averaged
 over several runs.  The performed simulations are listed in TABLE
 \ref{tab1} and TABLE \ref{tab2}.

We used the same $\tau$ definition as introduced in \cite{CsK}, namely:
\begin{equation}
\tau=\int\limits_0^\infty ({\rm min} (v^\ast (t),\langle v_\infty \rangle )
-\langle {\overline v}(t) \rangle ) dt 
\end{equation}
Where $v^\ast(t)$ is the velocity-time function of cars without interaction
and $v_\infty$ is the average velocity at the stationary state.
As in \cite{CsK} we found a peak in $\tau$ as a function of the density.
For the $v_{max}=2$ case $\tau_m$ the maximum of this peak is at 
$\rho_p\approx 0.15$
which is smaller
than $\rho_c=0.25$ [FIG.\ \ref{Fig1},FIG.\ \ref{Fig2}]. The reason of the 
appearing negative
relaxation times we see at FIG.\ \ref{Fig1} is that in this density region 
the average velocity
function $\overline{v}(t)$ has a maximum after a fast increase for short
times and it relaxes from this maximum to the stationary state [FIG.\ 
\ref{Fig3}].
Every timestep during this relaxation gives a negative contribution 
to $\tau$.
 
Assuming that
\begin{equation}
\tau_m(L) \sim L^{z} 
\end{equation}
we can fit $z\approx0.286 \pm 0.012$.  On the other hand $\sigma(L)$
the half-width of the $\tau$ peaks is roughly the same. Therefore, in
contrast to our earlier study with $v_{max}=5$, we have here no
implication to assume a
\begin{equation}
\sigma(L) \sim L^{-{1 \over \nu}}
\end{equation} 
scaling form and we could not do any finite size scaling. 

In the $v_{max}=1$ case $\rho_c=0.5$ as well known and $\rho_p\approx
0.35$ [FIG.\ \ref{Fig4}].  The curves seem to be more similar than in
the $v_{max}=2$ case and they seem to have the same form for different
system sizes .  The value of the $z$ exponent is $z=0.27\pm0.02$.
This value seems to be universal for all the $v_{max}$ parameters
taking notice of \cite{CsK} too.  However, in the $v_{max}=1$ case
this behavior is not bound to $\rho_p$ but it can be obtained in a
broad range of the density.

Looking for a different definition of a relaxation time we tried to fit a 
\begin{equation}
\label{xexp}
v(t)=v_\infty \left( 1-b*t^{-x}*\exp\left( -{t \over \tau}\right)
\right)
\end{equation}
function on the measured $\overline{v}(t)$ functions. Here $v_\infty$
is the average velocity in the stationary state and $b$, $x$ and $\tau$
are fitting parameters. We began the fit after the first increasing regime
which consists of $3$-$10$ steps because then the car movements are 
nearly independent and therefore  the $\overline{v}(t)$ graph
is roughly a straight line.
 
For every fit the value of the $\tau$ exponent arose as a greater value than
the corresponding system size. The exponent was vague to fit and had a 
great error because its effect falls in the regime of the fluctuations 
around $v_\infty(\rho)$ the stationary value of $\overline{v}(t)$.
In the $v_{max}=2$ case we could fit only in a restricted density region 
because of the shape of $\overline{v}(t)$ described above.

In the $v_{max}=1$ case the received values of the $x$ exponent were
the same within errors. Their value is $0.62\pm0.09$.
This value is near to $2/3$. This value can be interpreted in terms
of a mapping to the KPZ surface growth problem (see the next Section).

In the $v_{max}=2$ case the value of the $x$ exponent increases from
$0.6$ to $1.5$ in the measured density region [FIG.\ \ref{Fig5}]. We have 
an upper density limit for the fits because of the emerging maximum in
$\overline{v}(t)$ [FIG.\ \ref{Fig3}]. 

\section{Relation to Surface Growth}
\label{surface}
With an appropriate conversion this traffic model can be transformed 
into a surface growth model \cite{MR}. The transformation is as follows. 
We go along the lane from left to right and order a slant line to every 
lattice site whose length is $\sqrt{2}\times$lattice spacing and 
bevels with an angle $\alpha$ from the lane. At every step we order a line
with a slope of $\alpha=45^\circ$ to a car and a line
with a slope of $\alpha=-45^\circ$ to a hole. We always continue 
this emerging zig-zag line at the end of it and we make a step upward 
or downward according to whether we find a hole or a car. 
After every update of the traffic model we get a surface in this model.
The update of the traffic model determines the update of the surface
growth model [FIG.\ \ref{Fig6}].

If the density $\rho \neq 0.5$ the surface has a density dependent 
average steepness of $(2\rho-1)\sqrt{2}$ because the difference of the two ends 
of the surface is $\sqrt{2} (2\rho-1) L$. 
 
In the $v_{max}=1$ case the surface growth picture corresponds to a 
deposition model in which we drop squares with their corner downward 
in every local valley with probability 
$(1-p)$ as we go from left to right. 
The squares have an edge of 
$\sqrt{2}\times$lattice spacing.
The deposition of a square corresponds 
to a step with velocity $v=1$ in the traffic model.  

This simple deposition picture does not hold for the $v_{max}=2$ case because 
there are $v=2$ steps also which would correspond to the deposition 
of a rectangle or correlated deposition of squares. 

In our simulations we measured the time dependence of the average
surface width
\begin{equation} 
w^2(t)=\langle (h(i,t)-\overline{h}(t))^2 \rangle_i
\end{equation} 
where $h(i,t)$ is the height at site $i$, $\langle . \rangle_i$ means
averaging over the sites and $\overline{h}(t)$ is the average height function
$$\overline{h}=\langle h(i,t) \rangle_i \, .$$
In a broad class of growth models $w(t)$ behaves according to the 
Kardar-Parisi-Zhang theory \cite{BS} that is
\begin{equation}
w \sim t^\beta \qquad\qquad {\rm if} \qquad t\ll L^z
\end{equation} 
and
\begin{equation}
w\sim L^\alpha \qquad\qquad {\rm if} \qquad t\gg L^z \, .
\end{equation}
The simple deposition model corresponding to the $v_{max}=1$ case is
known to belong to this universality class \cite{MR}.

As initial condition we used equally distributed cars. For the
uniformly distributed initial condition we received large fluctuations
and have not seen any regularity.  In that case we saw the same
behavior as in the system with equally distributed initial condition
after relaxation.  We subtracted the average steepness from the
surface height in order to calculate $w^2(t)$.  The performed
simulations are listed in TABLE \ref{tab3} and TABLE \ref{tab4}.

In the $v_{max}=1$ case in the whole density region we see 
KPZ-like behavior [FIG.\ \ref{Fig7}] and  
we received $2\beta=0.63\pm0.04$ and $2\alpha=0.99\pm0.06$ which
should be compared to the KPZ values $\beta =1/3$ and $\alpha =1/2$.

This KPZ-like behavior explains the $2/3$-close value of the $x$
exponent received from the relaxation time analysis (\ref{xexp}) because
the velocity of the cars:
\begin{equation}
\overline{v}(t+1)={1 \over 2} \sum_{i=1}^L \left( (h(i,t+1)-h(i,t) \right)=
{L \over 2} (\overline{h}(t+1)-\overline{h}(t)) \ ;
\end{equation}
that is,
\begin{equation}
{2 \over L} \overline{v}(t)={d \overline{h} \over {dt} } \equiv v_s(L,t) \ . 
\end{equation}
According to \cite{KM}:
\begin{equation}
\Delta v_s(L,t)=v_s(L,t)-v_s^0 \sim t^{-\alpha_\perp} \qquad {\rm for} \qquad
t \ll L^z \ , 
\end{equation}
where $v_s^0$ is $v_s(L,t)$ for $t,L \rightarrow \infty$ . The exponent 
$\alpha_\perp$ corresponds to our $x$ exponent and is in one dimension
$x=\alpha_\perp=2/3$.

In the $v_{max}=2$ case the shape of the $w^2(t)$ curves is also KPZ-like.
We measured $2\alpha$ to be $2\alpha=1\pm0.05$. In contrast to the 
$v_{max}=1$ case the $\beta$ exponent depends on the density as visible
on [FIG.\ \ref{Fig8}] and it is not uniform as in the $v_{max}=1$ case.

\section{Correlation functions}
\label{corr}
Finally we studied the velocity correlation function in the space
of car series with the definition
\begin{equation}
C(i,t)=\left\langle \bigl( v(j,t^\prime) -{\overline v}(t^\prime) \bigr)
\bigl( v(j+i,t^\prime+t)-{\overline v}(t^\prime+t) \bigr) \right\rangle_j 
\end{equation}
or
\begin{equation}
\label{cfunct}
C^\prime(i,t)=\left\langle v(j,t^\prime) v(j+i,t^\prime+t) \right\rangle_j
\end{equation} 
where $v(j,t)$ is the velocity of the $j$-th car at time $t$ and $\langle
.\rangle_j $ means averaging over all cars and $t^\prime$ is a time when
the system is in the steady state. Being in the stationary state $C(i,t)$
is independent of $t^\prime$.

If we consider the function $C(i,t)$ at a constant time $t$ as a function of
$i$, we find a peak centered at the value
\begin{equation}
i(t)=\max_j C(j,t) \, .
\end{equation}
The maximum of the peak decreases with increasing $t$ [FIG.\ \ref{Fig9}]. 
$i(0)=0$ because 
that is the autocorrelation function but for greater times it appears at other
car indexes. Considering the $i(t)$ versus $t$ graph we can fit a straight line
on the $i(t)$ values which lines steepness depends on the density.
This steepness defines a velocity in the car series space with which the peak 
spreads.

Except for the peak the $C(i,t)$ function fluctuates around a constant value
and it forms a plateau.

From the definition of $C(i,t)$ it follows that the jammed cars do not
contribute to $C(i,t)$ since they are staying. The cars already
accelerated near $v_{max}$ give contribution to the plateau of $C(i,t)$. 
Those cars give contribution to the peak which accelerate in a correlated
manner because a car accelerating makes place to the car behind to 
accelerate. These are the cars coming out of a jam.

Therefore the longterm existence of the peak and its velocity characterizes
the traffic jams in the system.
 
In the $v_{max}=1$ case the $V_1(\rho)$ velocity of the peak increases without
any sign of criticality from zero to its value at $\rho=1$ where 
$v(\rho \to 1)=(1-p)$. This value follows from the motion of one single 
hole [FIG.\ \ref{Fig10}].

However, in the $v_{max}=2$ case we can see a bending at the
$V_2(\rho)$ function at the density $\rho_k=0.125$ where the function
shows a steep increase [FIG.\ \ref{Fig11}].  This implies emerging traffic
jams. The small $V_2(\rho)$ values of the function below $\rho_k$ come
from the small fugitive jams being in the system.  It is tempting to
interpret the bending in $V_2(\rho)$ as the appearance of a new phase
at $\rho_k$, since the behavior of the peak velocity is reminiscent to
an order parameter of second order phase transitions.  However, the
velocity does not go to zero at the bending point and -- in contrast
to what is expected for second order transitions -- we could not
observe any finite size scaling.

\section{The $v_{max}=\infty$ case}
\label{vinf}
Our simulations have indicated the following: \hfill\break 
i) The behavior
$v_{max}>1$ is different from $v_{max}=1$ \hfill\break 
ii) For
$v_{max}>1$ there is a jamming transition but a careful study of the
different characteristics indicated -- in contrast to what was
suggested in \cite{CsK} -- that there is no clear second order phase
transition related.  \hfill\break 
iii) From the present and earlier studies \cite{NP,CsK,NH}
it is clear that the smaller $p$ the
more pronounced is the transition. \hfill\break  
iv) Comparing our results for $v_{max}=2$ with those with
$v_{max}=5$ we realize that increasing $v_{max}$ also sharpens the
transition. 

The origin of iii) is clear: In the $p=0$ model there is a phase
transition related to a singularity in the fundamental diagram.  In
order to see the origin of iv) we considered the case without speed
limit, more precisely the $v_{max}=L$ case.
 
The fundamental diagram of the $v_{max}=\infty$ case shows different
features than the cases discussed so far [FIG.\ \ref{Fig12}]. The main
characteristics of its shape is a plateau value depending on $p$
[FIG.\ \ref{Fig13}]. The function decreases fast to this plateau. The height of
the plateau is independent of the system size only its length changes
with.

Looking at the flow of the cars we can observe that at the densities
where the current of the stationary state $j_\infty(\rho)$ is higher
than the $j_p$ plateau value the cars tend to be equally distributed
what is typical for the deterministic case \cite{SSNI}. However, at the
density regime of the plateau we see one jam which characterizes this
density region [FIG.\ \ref{Fig14}]. For higher densities where
$j_\infty(\rho)$ decreases with the density we see more than one jam.

Calculating the (\ref{cfunct}) correlation function we can notice that the
$v_\infty(\rho)$ peak velocity in the plateau region corresponds to
$j_p$ which also means that we have only one jam in the system for
this densities.
  
Assuming that the $\rho_t$ value where the plateau sets in depends on
$L$ like
\begin{equation}
\rho_t \sim L^{-t}
\end{equation}
we got $t\approx 0.5$. Because this point represents the $N_f=\rho_t
L$ number of cars being in the flow between the right and left side of
the jam we can conclude that the density contribution of cars in the
flow part scales as
\begin{equation}
\rho_f \sim L^{-t}
\end{equation}
and therefore the density contribution of the cars being in the jam
will be
\begin{equation}
\rho_j=\rho-\rho_f
\end{equation}
\begin{equation}
\rho_j \to \rho \qquad {\rm when} \qquad L \to \infty.
\end{equation}
Thus values greater than the $j_p$ plateau value of $j_\infty(\rho)$
(except the value at $\rho=0$) can be considered as fluctuation
effects. When $\rho \to 0$ we have finite number of cars with infinite
velocity (one car with velocity $L-1$).  In the thermodynamic limit
\begin{equation}
j_\infty(\rho \to 0)={L-1 \over L} \qquad {\rm as} \qquad L \to \infty.
\end{equation} 

Therefore we can say that in the thermodynamic limit we have a phase transition
at $\rho=0$ between the phases of $j_\infty(0)=1$ and no jams and of the 
$j_{p}$ value with one jam. Increasing $\rho$ the plateau ceases and there
are more than one jams in the system.

This result contradicts the result of the mean field theory described
in \cite{SSNI} where the fundamental diagram starts in the $\rho=0$
$j=0$ point with infinite slope.

For smaller braking probabilities than $p=0.001$ the plateau ceases and only
a break in the graph remains.
 
\section{Conclusions}

According to our comparison of the three different models we can
conclude to the following.  As we summarized at the beginning of
Section 6 there is a difference between the behavior of the models
$v_{max}=1$ and $v_{max}>1$. In contrast with the $v_{max}=1$ model in
the $v_{max}=2$ case we see a kind of transition but it is not a
strict second order transition it seems to be a crossover
transition. We found that the density at which the transition takes
place is smaller then the $\rho_c=0.19$ value mentioned in \cite{SS}.
On the grounds of the obtained results for $v_{max}=\infty$ where we
found a first order transition we can conclude the the increase of the
$v_{max}$ parameter makes the transition more striking.  Our results
indicate that the behavior of the models labeled by different $p$ and
$v_{max}$ parameters is guided by the transition points of the models
with parameters $p=0$ or $v_{max}=\infty$.

\begin{table}
\caption{Performed simulations for the relaxation time analysis for 
$v_{max}=1$} 
\begin{tabular}{lr}
System size & Number of runs \\
\tableline
$L=2000$ & $10000$ \\
$L=4000$ & $5000$ \\
$L=8000$ & $2000$ \\
\end{tabular}
\label{tab1}
\end{table}

\begin{table}
\caption{Performed simulations for the relaxation time analysis for
$v_{max}=2$ }
\begin{tabular}{lr}
  System size & Number of runs  \\
\tableline
 $L=2048$ & $10000$  \\
 $L=4096$ & $5000$  \\
 $L=16384$ & $1000$  \\
 $L=32768$ & $500$  \\
\end{tabular}
\label{tab2}
\end{table}

\begin{table}
\caption{Performed simulations for the measuring of $w^2(t)$ for $v_{max}=1$ }
\begin{tabular}{lr}
System size & Number of runs \\
\tableline
$L=1024$ & $500$ \\
$L=2048$ & $200$ \\
$L=4096$ & $50$ \\
\end{tabular}
\label{tab3}
\end{table}

\begin{table}
\caption{Performed simulations for the measuring of $w^2(t)$ for $v_{max}=2$ }
\begin{tabular}{lr}
System size & Number of runs \\
\tableline
$L=512$ & $20000$ \\
$L=1024$ & $10000$ \\
$L=2048$ & $5000$ \\
\end{tabular}
\label{tab4}
\end{table}


\begin{figure}
\caption{
Relaxation times versus density measured for the $v_{max}=2$
case for different system sizes. The maximum of the peaks is at the density
$\rho_p \approx 0.15$ and they have nearly the same halfwidths.
}
\label{Fig1}
\end{figure}

\begin{figure}
\caption{
Fundamental diagram for the $v_{max}=2$ model measured for
$L=2048$. The maximum of the diagram is at $\rho_c=0.25$.
}
\label{Fig2}
\end{figure}

\begin{figure}
\caption{
Average velocity versus time graph for $v_{max}=2$,
$L=2048$, $\rho=0.3125$.
}
\label{Fig3}
\end{figure}

\begin{figure}
\caption{ Relaxation times versus density graph measured for $v_{max}=1$.
We see similar curves for different system sizes.  }
\label{Fig4}
\end{figure}

\begin{figure}
\caption{ The $x$ exponent versus density graph for $v_{max}=2$.  }
\label{Fig5}
\end{figure}

\begin{figure}
\caption{ Determination of the corresponding surface growth model
to the traffic model.  }
\label{Fig6}
\end{figure}

\begin{figure}
\caption{ The square of the average surface widths versus time graph
for the $v_{max}=1$ case for different densities. The system size is
$L=2048$ and the curves correspond to the following densities.
$\rho=0.125$ ($N=256$), $\rho=0.25$ ($N=512$), $\rho=0.5$ ($N=1024$),
$\rho=0.75$ ($N=1536$), $\rho=0.96875$ ($N=1984$).
On the log-log plot the the KPZ-like behavior is apparent.  }
\label{Fig7}
\end{figure}

\begin{figure}
\caption{ $2\beta$ exponent measured in the $v_{max}=2$ model for different
densities. }
\label{Fig8}
\end{figure}
 
\begin{figure}
\caption{ Correlation times versus car index for different times.
The concerned system is $L=256$, $\rho=0.5$, $v_{max}=2$. }
\label{Fig9}
\end{figure}

\begin{figure}
\caption{ The peak velocity $V_1(\rho)$ as a function of the density
for $v_{max}=1$.  }
\label{Fig10}
\end{figure}

\begin{figure}
\caption{ The peak velocity $V_2(\rho)$ as a function of the density
for $v_{max}=2$.  }
\label{Fig11}
\end{figure}

\begin{figure}
\caption{ Fundamental diagram for $v_{max}=\infty$ for $L=1000$. }
\label{Fig12}
\end{figure}

\begin{figure}
\caption{ The plateau values for different $p$ braking probabilities
in the $v_{max}=\infty$ model.  }
\label{Fig13}
\end{figure}

\begin{figure}
\caption{ The one-jam phase of the $v_{max}=\infty$ model. 
The horizontal direction means the lattice sites the vertical direction
the consecutive timesteps.  }
\label{Fig14}
\end{figure}

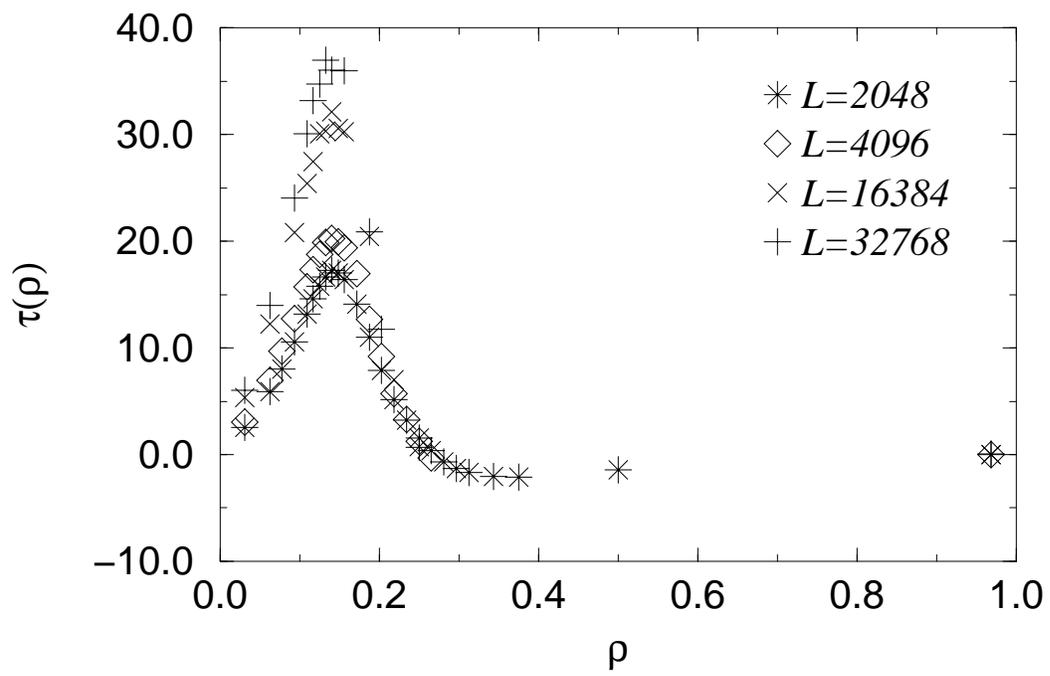
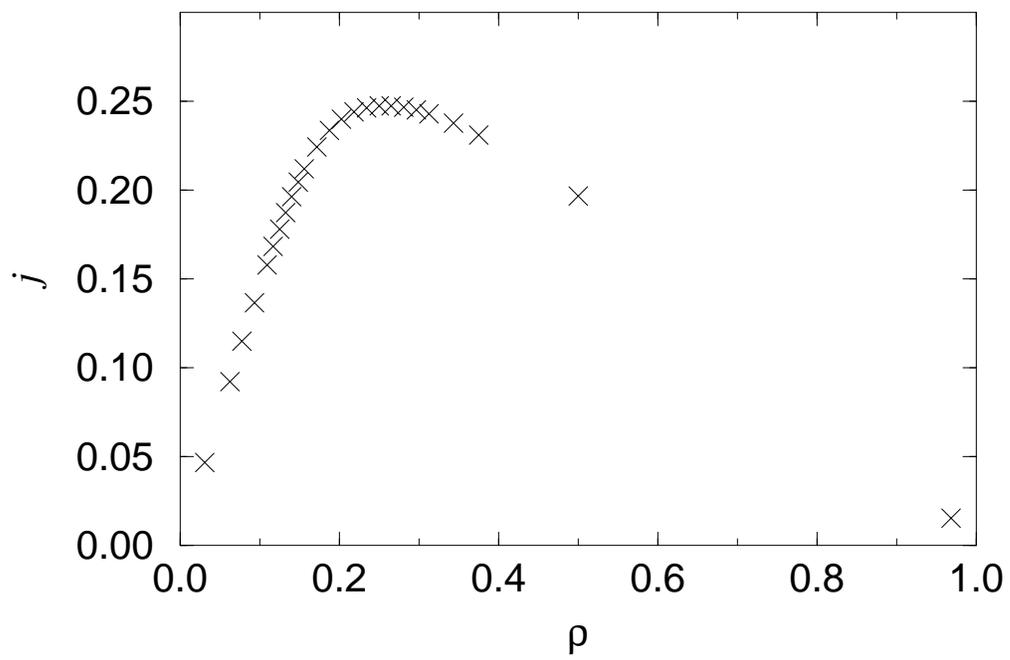
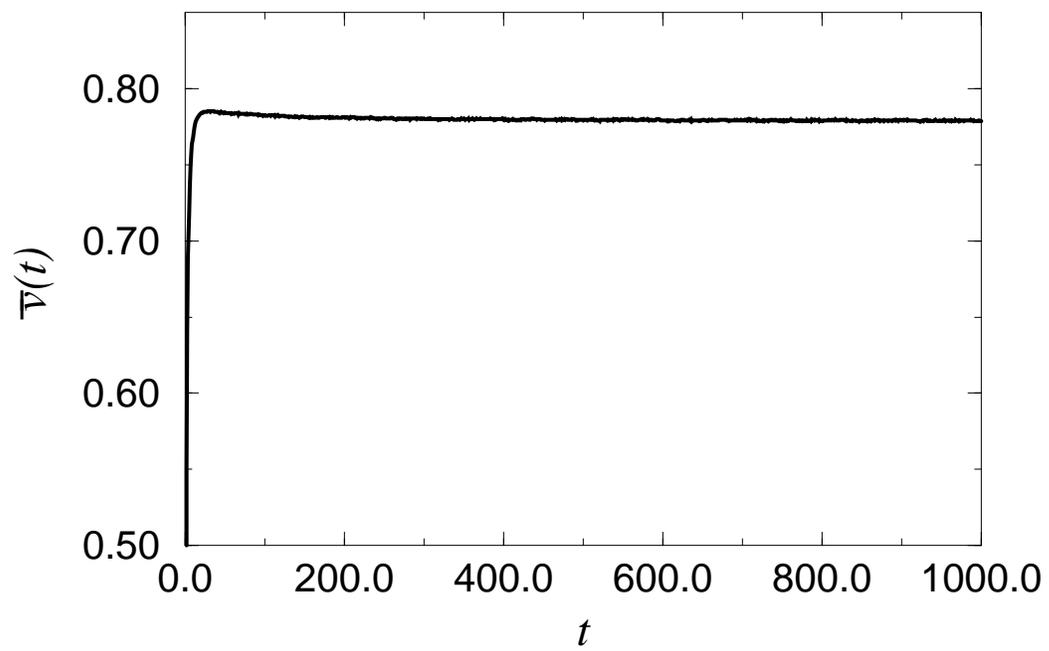
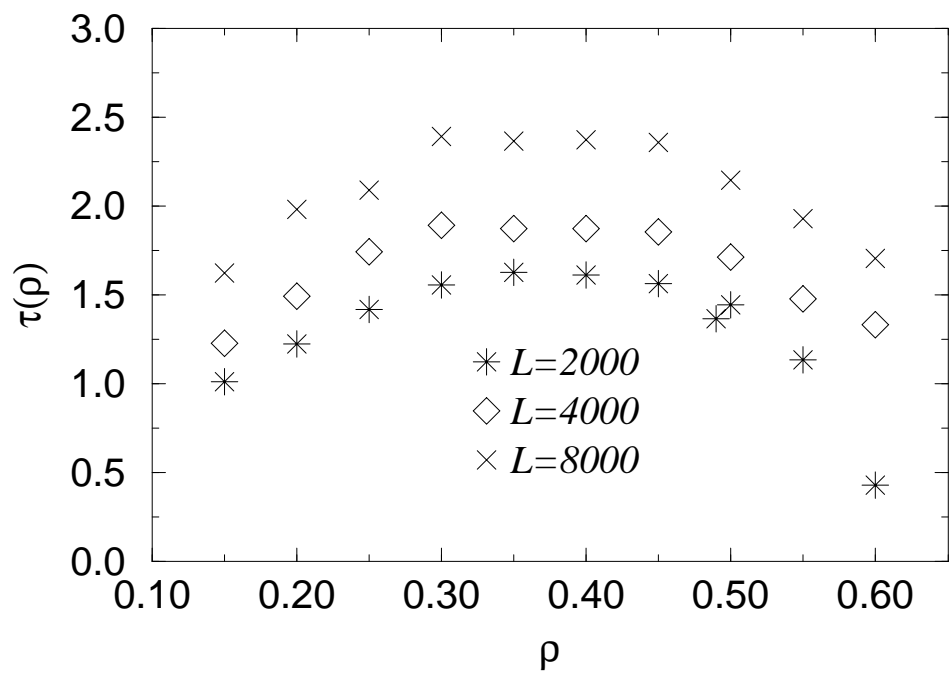
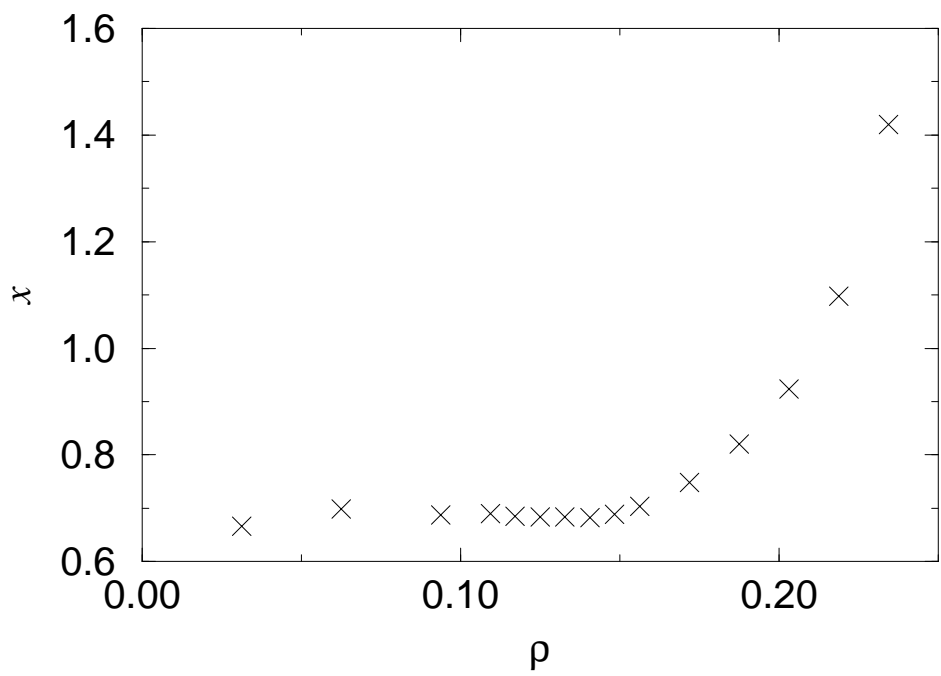
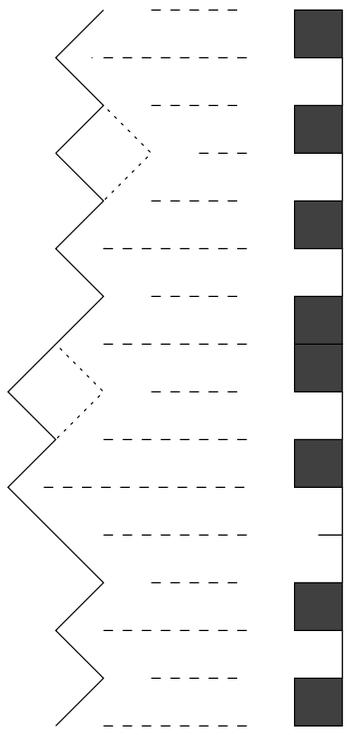
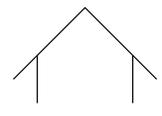
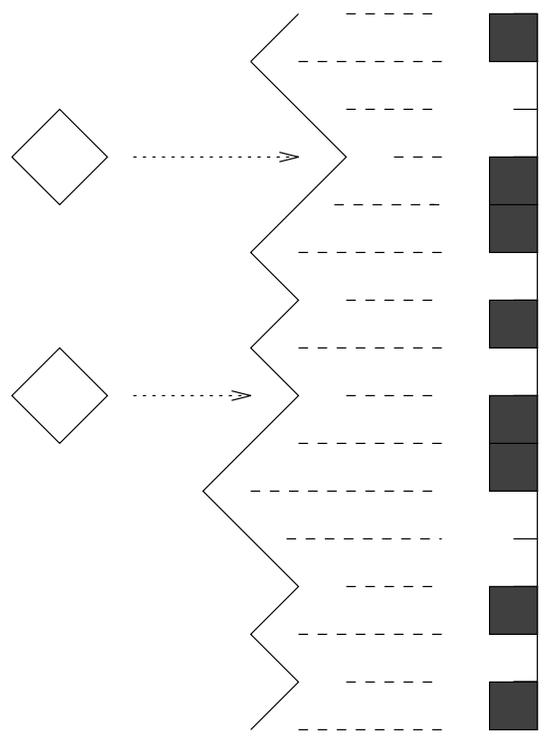
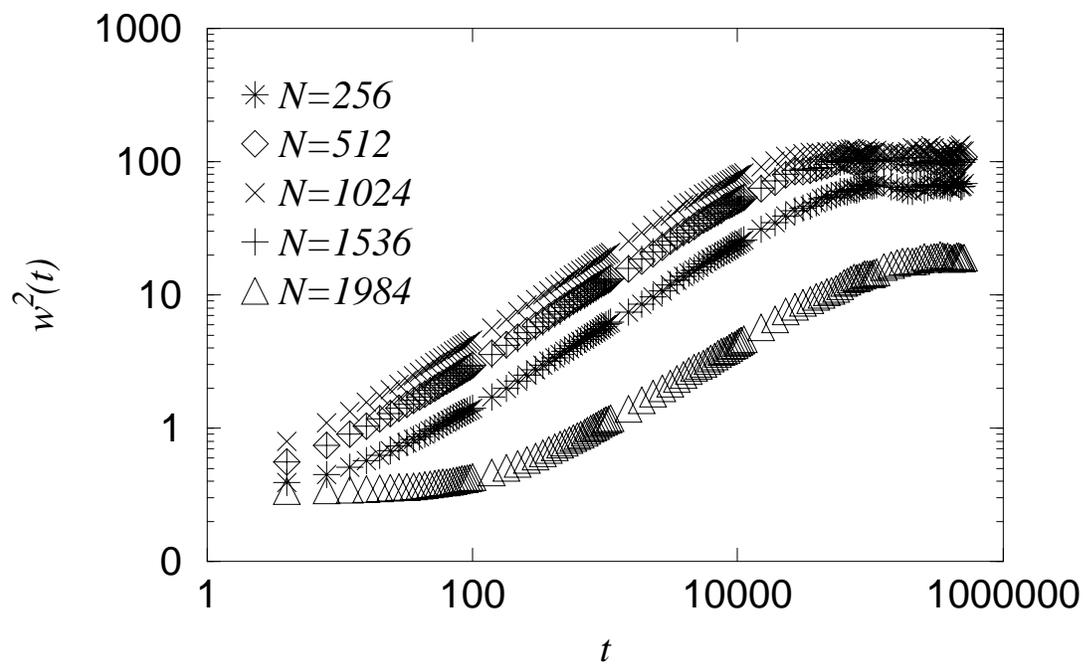
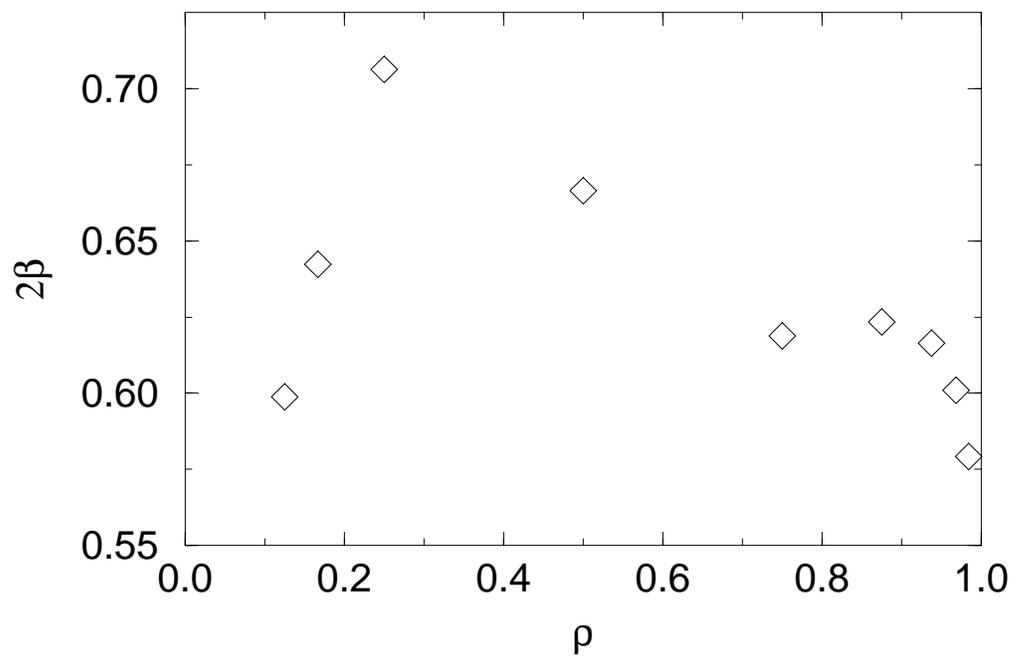
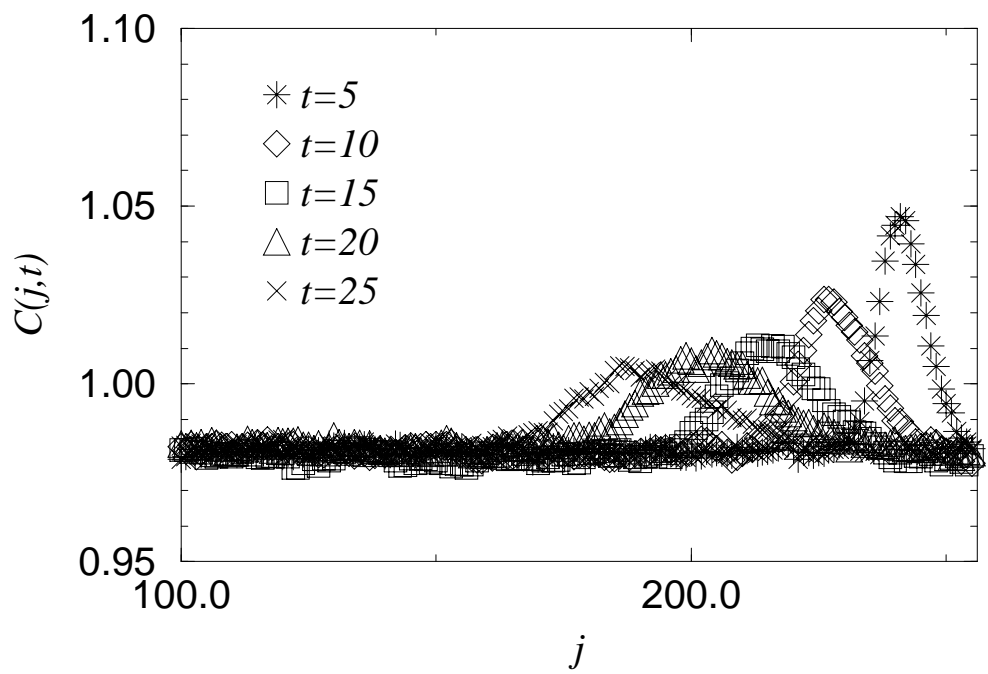
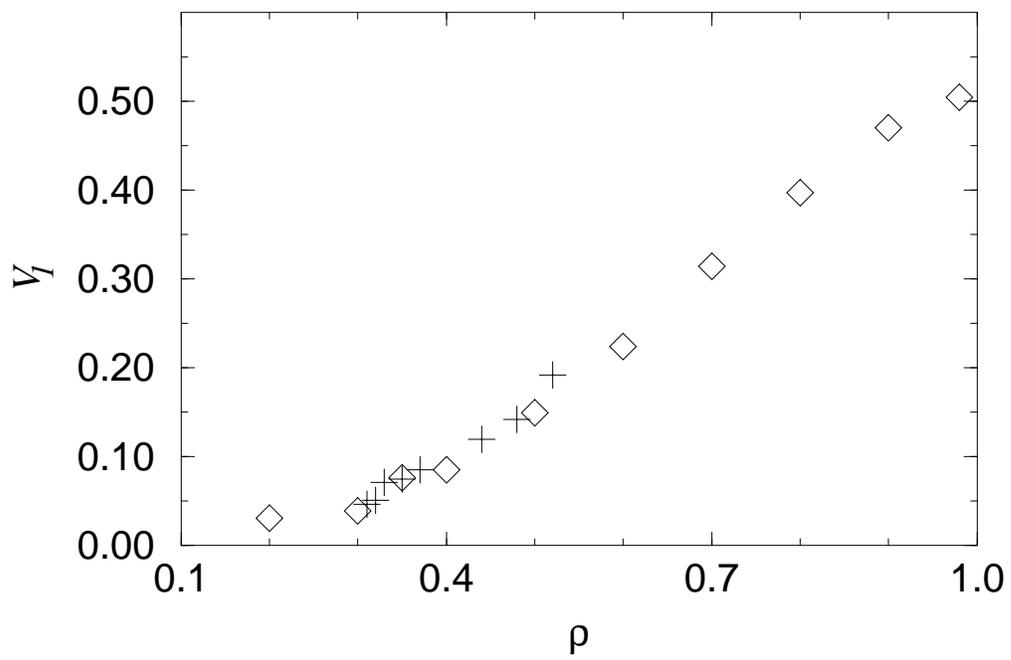
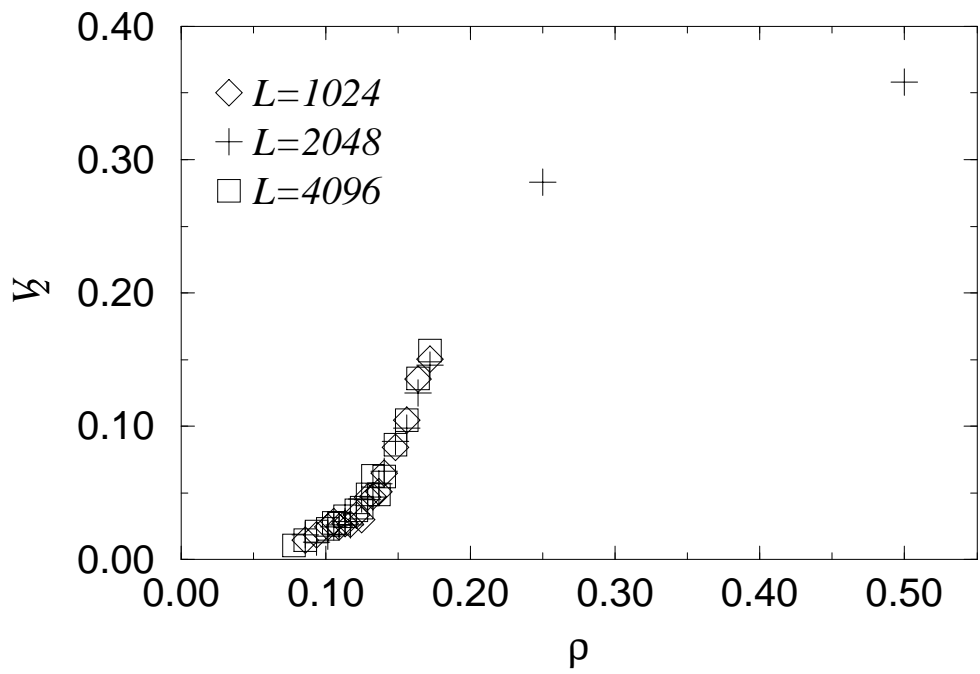
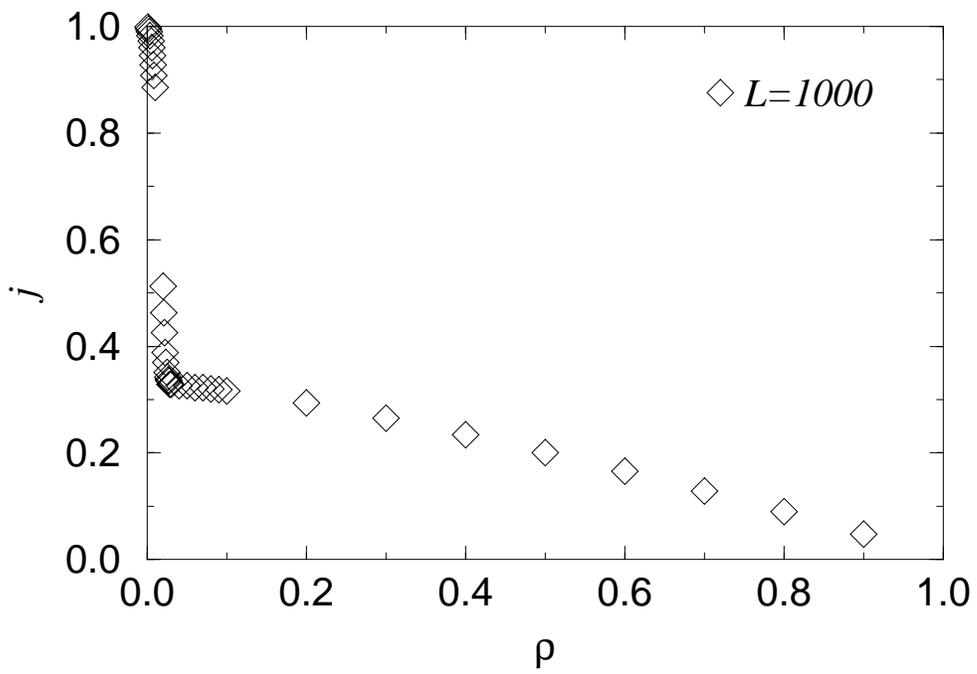
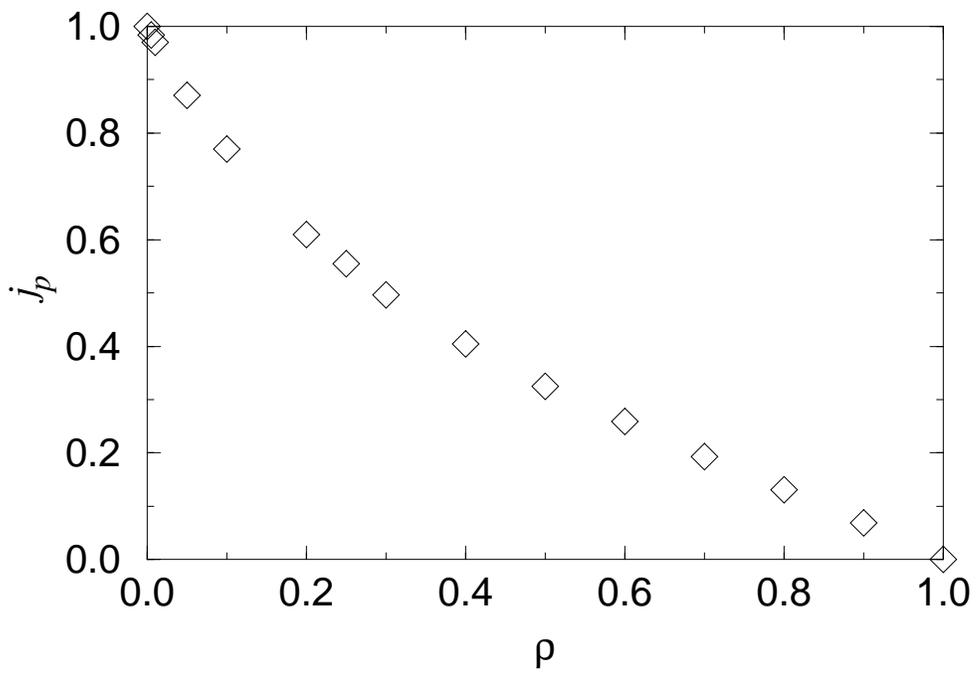
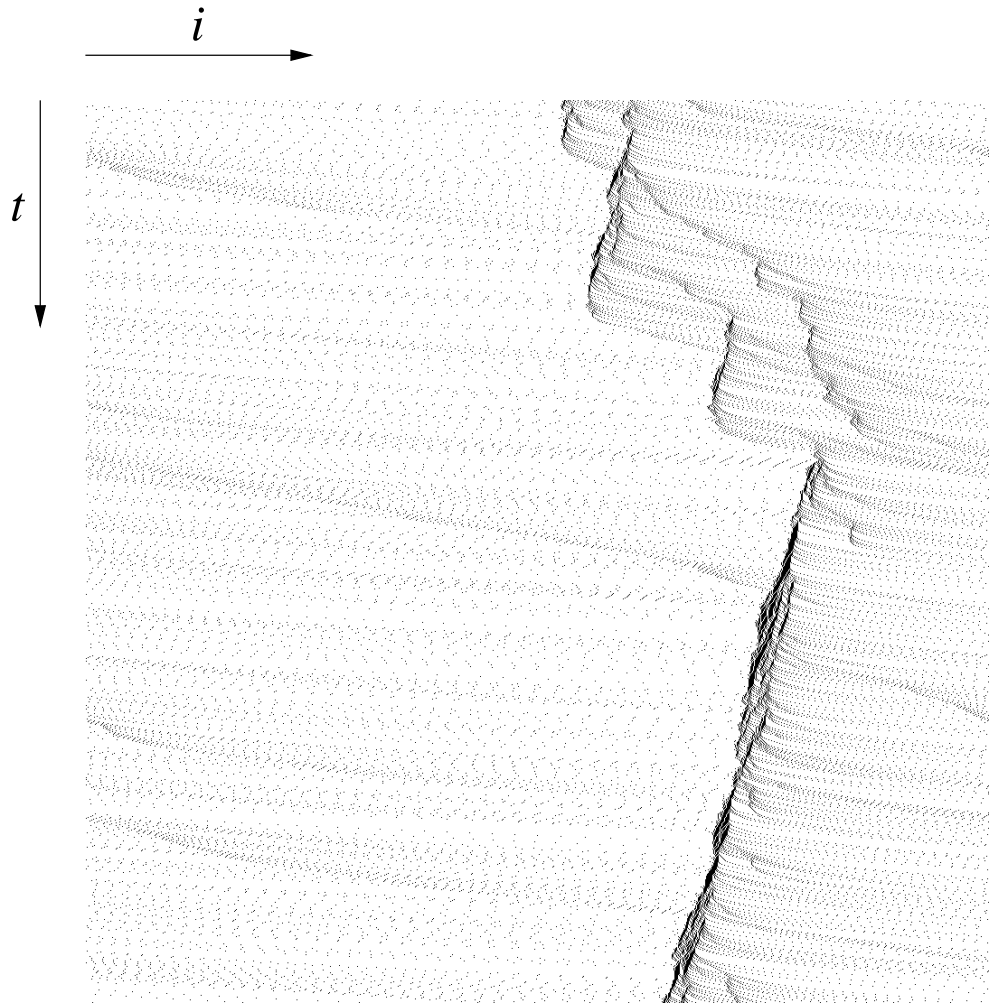
\end{document}